\documentclass[3p,times,procedia]{elsarticle}
\usepackage{nupha_ecrc}

\volume{00}

\firstpage{1}

\journalname{Nuclear Physics A}

\runauth{}

\jid{nupha}

\jnltitlelogo{Nuclear Physics A}

\usepackage{amssymb}

\usepackage[figuresright]{rotating}
\usepackage{subfigure}
\usepackage{bm}
\usepackage{xcolor}

\newcommand{\Raa}{$R_{\rm{AA}}$}
\newcommand{\ptjet}{$p_{\rm{T,jet}}$}
\newcommand{\pt}{$p_{\rm{T}}$}
\newcommand{\Rpar}{$R$}
\newcommand{\snn}{$\sqrt{s_{\rm{NN}}}$}

\begin{document}

\begin{frontmatter}



\dochead{XXVIIIth International Conference on Ultrarelativistic Nucleus-Nucleus Collisions\\ (Quark Matter 2019)}

\title{Quenching of heavy and light flavour jets: experimental overview}


\author{Barbara Trzeciak}

\address{Faculty of Nuclear Sciences and Physical Engineering \\Czech Technical University in Prague\\ Brehova 7, 115 19 Prague, Czech Republic}

\begin{abstract}

In these proceedings, recent experimental results on jet quenching are discussed, focusing on measurements shown at the Quark Matter 2019 conference.

\end{abstract}

\begin{keyword}
heavy-ion collisions \sep jets \sep quenching \sep heavy-flavour \sep LHC \sep RHIC


\end{keyword}

\end{frontmatter}


\section{Introduction}
\label{intro}

In the ultrarelativistic high-energy nuclear collisions, partons originating from the initial hard scatterings serve as probes of the structure of the medium created in the collisions. While traversing the hot and dense medium (QGP), the hard-scattered partons interact with the medium and lose part of their energy. It leads to an effect known as the \textit{jet quenching}. The details of the parton-medium interactions and the energy loss mechanism are however still not fully understood. The thermodynamical and transport properties of the QGP, such as the medium transport coefficient $\hat{q}$, can be inferred from a model-to-data comparison.  

Experimentally, the jet quenching can be studied via a modification of the high-\pt\ particle production or of fully reconstructed jets that are defined as collimated sprays of hadrons with a given resolution parameter \Rpar, most commonly reconstructed using the anti--$k_{\rm T}$ algorithm~\cite{Cacciari:2008gn}. 
One can access the expected flavour dependence of the quenching for example by tagging jets with photons (thus increasing quark-seeded jet contribution) or with heavy-flavour hadrons. In central AA collisions, following has been observed: (i) a suppression of high-\pt\ charged hadrons, similar for different collision energies both at RHIC and LHC~\cite{Khachatryan:2016odn} (ii) no evidence of a flavour dependence at high \pt\ $>$~10~GeV/$c$ for charm ~\cite{Acharya:2018hre} and $>$~20~GeV/$c$ for beauty~\cite{Sirunyan:2018ktu}, (iii) a strong suppression of the fully reconstructed jets up to \pt\ of a TeV~\cite{Aaboud:2018twu}.  

In these proceedings, the recent experimental results on heavy- and light-jet quenching are discussed, focusing on a selection of measurements shown at this conference. The observables, related to the jet substructure, are excluded as they are a subject of a separate contribution.

\section{Nuclear Modification Factor}
\label{Raa}

The nuclear modification factor, \Raa, quantifies the deviation of inclusive spectra from the baseline, and is defined as: 
\begin{equation}
R_{\mathrm{AA}}=\frac{\frac{1}{N_{\mathrm{event}}} \mathrm{d} \mathrm{N}^{\mathrm{AA}} / \mathrm{d} p_{\mathrm{T}} \mathrm{d} \eta}{<T_{\mathrm{AA}}>\mathrm{d} \sigma^{\mathrm{pp}} / \mathrm{dp}_{\mathrm{T}} \mathrm{d} \eta},
\end{equation}
where $N^{\rm {AA}}$ is the yield in AA collisions, $\sigma^{\rm {pp}}$ is the proton-proton baseline cross section, and the $T_{\rm {AA}}$ is the nuclear overlap function. 

\subsection{$R$ scan}

Studying jets with different resolution parameters \Rpar\ gives an insight into the interplay of the suppression and recovery of the quenched energy. Systematic analyses of the jet nuclear modification factor as a function of the jet \pt\ and \Rpar\ provide additional constrains on the jet quenching models, and are complementary to jet substructure studies.

\begin{figure}[!htb]
\centering
 \subfigure[~]{\includegraphics[width=0.51\textwidth]{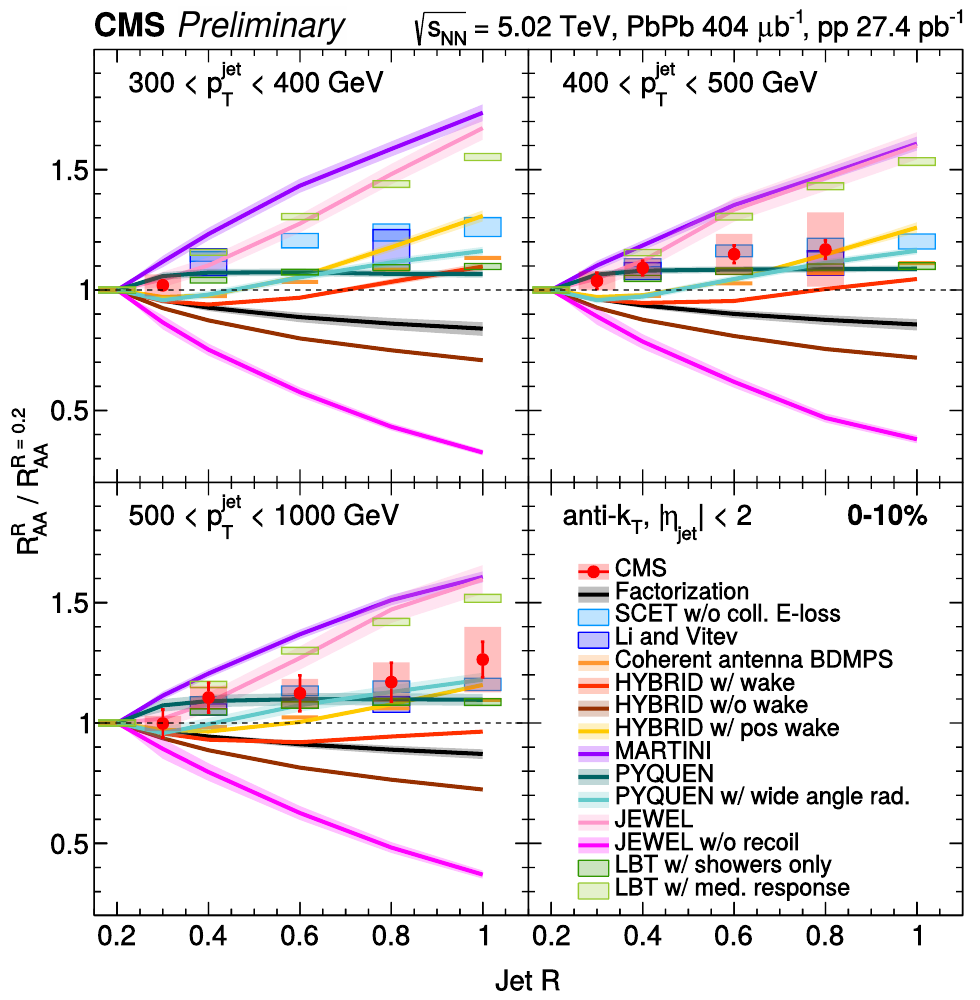}}
  \subfigure[~]{\includegraphics[width=0.47\textwidth]{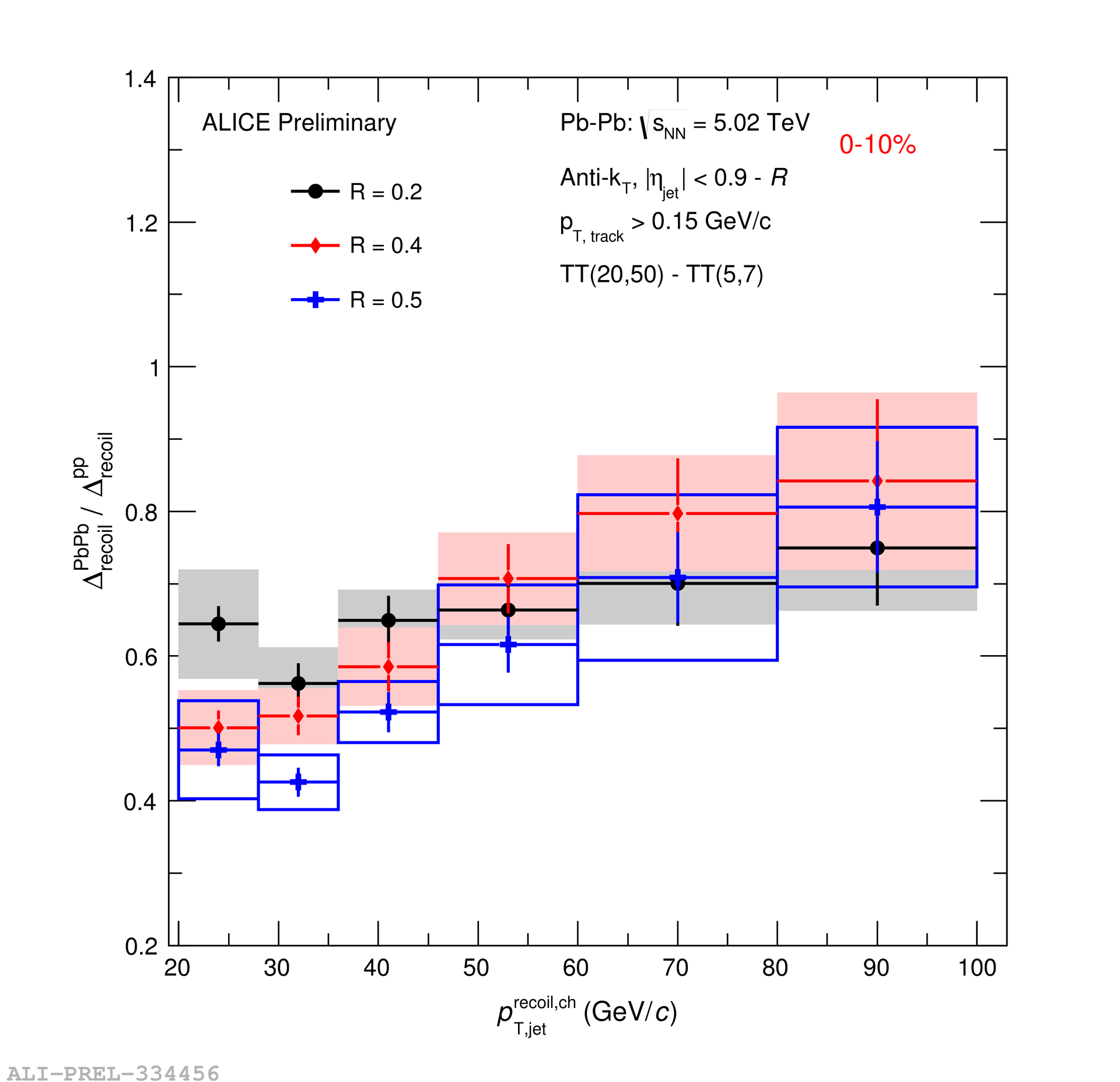}}\\
\caption{(a) Jet \Raa\ ratio to \Raa (\Rpar\ = 0.2) as a function of the \Rpar\ (in the range 0.3--1) for \ptjet\ ~$>$~300~GeV/$c$, in 0--10\% central Pb--Pb collisions at $\sqrt{s_{\rm{NN}}}$~=~5.02~TeV measured by the  CMS experiment~\cite{CMS:2019btm}
(b) Pb--Pb/pp $\Delta_{\rm {recoil}}$ ratio in most central Pb--Pb collisions at \snn ~=~5.02~TeV for \Rpar\ = 0.2, 0.4 and 0.5 measured by ALICE~\cite{YMao_theseproceedings}.}
\label{fig:RaaVsR}
\end{figure}

Figure~\ref{fig:RaaVsR}(a) presents jet \Raa\, normalised to \Raa\ (\Rpar\ ~=~0.2), for several \Rpar\ parameter values between 0.2 and 1 measured by the CMS experiment, for \ptjet~$ > $~300 GeV/$c$ in 0--10\% central Pb--Pb collisions at \snn\ ~=~5.02~TeV~\cite{CMS:2019btm}. At high \ptjet, the data show a mild increase of the $R_{\rm{AA}}^{R}/R_{\rm{AA}}^{R=0.2}$ ratio as a function of the jet~R.
The ALICE collaboration performed \Raa\ measurements at a lower jet \pt\ ranges, namely below 200 GeV/$c$ down to \ptjet\ of 40--60~GeV/$c$, for the \Rpar\ parameter values of 0.2 and 0.4~\cite{Acharya:2019jyg} and recently also for R~=~0.6~\cite{Haake:2019pqd}. 
A strong jet suppression with a very little \Rpar\ dependence was observed.

The experimental results were compared to various predictions of jet quenching theoretical models, event generators and analytical calculations~\cite{Khachatryan:2015pea,Zapp:2013vla,Pablos:2019ngg,He:2018xjv,Schenke:2009gb,Qiu:2019sfj,Baier:1994bd,Chien:2015hda,Li:2018xuv}. 
In the case of models, the different treatment of how the quenched energy is radiated and dissipated in the medium and of the medium response can result in significantly different trends of the jet \Raa\ vs \Rpar , especially at large \Rpar . This shows that the \textit{basic} \Raa\ observable when extracted $vs$ $R$ and \ptjet\ has a strong constrain power among different jet quenching models. And data covering wider \ptjet\ and $R$ ranges, in different collision centrality intervals, would be of a great importance here.

The current experimental results are limited to rather high jet \pt\ and/or small \Rpar, in particular in the most central collisions. This is due to the large and fluctuating background of the underlying event of particles uncorrelated with the hard scattering. At low \pt\ and in central collisions stronger medium effects on the jet are however expected. One of a promising solutions to deal with the large background corrections at low jet \pt\ and large \Rpar\ is to utilize Machine Learning techniques~\cite{Haake:2018hqn}, as was demonstrated by the ALICE experiment~\cite{Haake:2019pqd}.

\subsection{Semi-inclusive jet measurements}

Another way of performing jet analysis at lower \pt\ and large \Rpar\ parameter is to measure semi-inclusive jets recoiling from high-\pt\ trigger hadrons.
This approach minimises jet selection biases. In addition, the large uncorrelated background can be suppressed using data-driven techniques. 

The ALICE collaboration measured the $\Delta_{\rm {recoil}}$ observable, defined as a difference between two semi-inclusive recoil jet distributions:
\begin{equation}
\Delta_{\rm {recoil }}= \left. \frac{1}{N_{\rm {trig }}} \frac{\mathrm{d}^{2} N_{\rm {jet }}}{\mathrm{d} p_{\rm {T, jet }}^{\mathrm{ch}} \mathrm{d} \eta_{\rm {jet }} } \right|_{p_{\rm {T.trig }} \in \mathrm{TT}_{\rm {Sig }}}
-\left.c_{\rm {Ref }} \cdot \frac{1}{N_{\rm {trig }}} \frac{\mathrm{d}^{2} N_{\rm {jet }}}{\mathrm{d} p_{\rm {T,jet }}^{\rm {ch }} \mathrm{d} \eta_{\rm {jet }}}\right|_{p_{\rm {T.trig }} \in \mathrm{TT}_{\rm {Ref }}}
\end{equation}
for \Rpar\ ~=~0.2, 0.4 and 0.5 at 20~$< p_{\rm{T,jet}} < $~100~GeV/$c$~\cite{YMao_theseproceedings}.
Figure~\ref{fig:RaaVsR}(b) shows $\Delta_{\rm {recoil }}^{\rm{PbPb}} / \Delta_{\rm {recoil }}^{pp}$ in the most central Pb--Pb collisions at \snn\ ~=~5.02~TeV. 
The results for different $R$ are consistent within the uncertainties with a hint of increasing suppression with increasing $R$ values at \pt\ below 40~GeV/$c$.

In the STAR experiment at RHIC, per-trigger recoil jet yields, $Y(p_{\rm{T,jet}})$, triggered with $\pi^{0}$ ($11<\mathrm{E}_{\mathrm{T}}^{\rm {trig }}<15$ GeV/$c$) were studied for two \Rpar\ parameter values of 0.2 and 0.5~\cite{NSahoo_poster}. At \snn\ = 200~GeV, the \Rpar\ ~=~0.2 results show jet suppression in central Au--Au collisions, while at \Rpar\ ~=~0.5 an indication of the energy recovery is observed, $Y^{\rm{AuAu}}/Y^{\rm{pp}}$ consistent with unity within the uncertainties with Pythia8 predictions used as the baseline.

\section {Path-length dependence of the energy loss}

The path-length dependence of the parton in-medium energy loss can be probed by performing collision-system dependent measurements of the high-\pt\ hadron yield suppression. The LHC provided Pb (\snn ~=~5.02 TeV) and Xe (\snn ~=~5.44 TeV) ion beams. Taking into account the sizes of the two nuclei and under an assumption of quadratic (linear) dependence of the energy loss on the traversed path length, one expects a 31\%(17\%) reduction of the lost energy in head-on Xe--Xe collisions w.r.t. Pb--Pb~\cite{Sirunyan:2018eqi}. 
Indeed, both CMS~\cite{Sirunyan:2018eqi} and ALICE~\cite{Acharya:2018eaq} reported $R_{AA}$ in Xe--Xe to be larger compared to the Pb--Pb results in the same centrality classes. However, the data do not have enough constraining power among different models. An observable more sensitive to the path length dependence of the parton energy loss was proposed: $R_{L}^{XePb} = (1- R_{AA}^{XeXe})/(1-R_{AA}^{PbPb})$, which may directly differentiate between different energy-loss mechanism from the experimental data~\cite{Djordjevic:2018ita}. The LHC data, together with the recent PHENIX measurements~\cite{PHENIX_AHodges_theseProceedings} of the $R_{AA}$ in U--U, Au--Au and Cu--Cu collisions, also show that the high-\pt\ $R_{AA}$ in different systems is compatible with each other at the same $N_{part}$ value, which further proves dependence on the system size rather than the collision species.

STAR experiment introduced a technique called Jet Geometry Engineering (JGE) which aims to control the sampled parton path length in medium by biasing the selected jet sample~\cite{Adamczyk:2016fqm, Raghav_theseProceedings}.
At RHIC energies, a requirement on a high momentum hadron trigger in a jet selects jets that are closer to the interaction region surface~\cite{Renk:2012cx}. In addition, one can analyse di-jet configurations and by applying higher threshold on the momentum of the jet constituents select recoil jets with even longer path length in the medium. The main idea behind the JGE studies is to vary the constituent threshold and the jet radius in order to select di-jet samples with potentially different path lengths and so different energy losses. However, a theoretical input is essential to constrain the path length dependence of the partonic energy loss from these studies.

\section{Redistribution of the quenched energy}

It was observed that particles with low momenta associated with jets have a much broader angular distribution extending outside the jet~\cite{Chatrchyan:2011sx,Khachatryan:2015lha,Khachatryan:2016tfj,Sirunyan:2018jqr,Acharya:2019ssy} and their production is enhanced as compared to $pp$~case \cite{Aad:2014wha,Chatrchyan:2014ava,Aaboud:2017bzv,Aaboud:2018hpb}. It suggests that the quenched energy is redistributed to large angles and in multiplicities of softer particles via a soft gluon emission.
The ATLAS collaboration measured a distribution of charged particles at a distance $r$ around the axis of jets with \Rpar\ ~=~ 0.4~\cite{Aad:2019igg}:
\begin{equation}
D(p_{\rm T},r) = \frac{1}{N_{\rm{jet}}} \frac{1}{2 \pi r dr}  \frac{dn_{\rm{ch}}(p_{\rm T},r)}{dp_{\rm T}}
\end{equation}

\begin{figure}[!htb]
\centering
 \subfigure[~]{\includegraphics[width=0.45\textwidth]{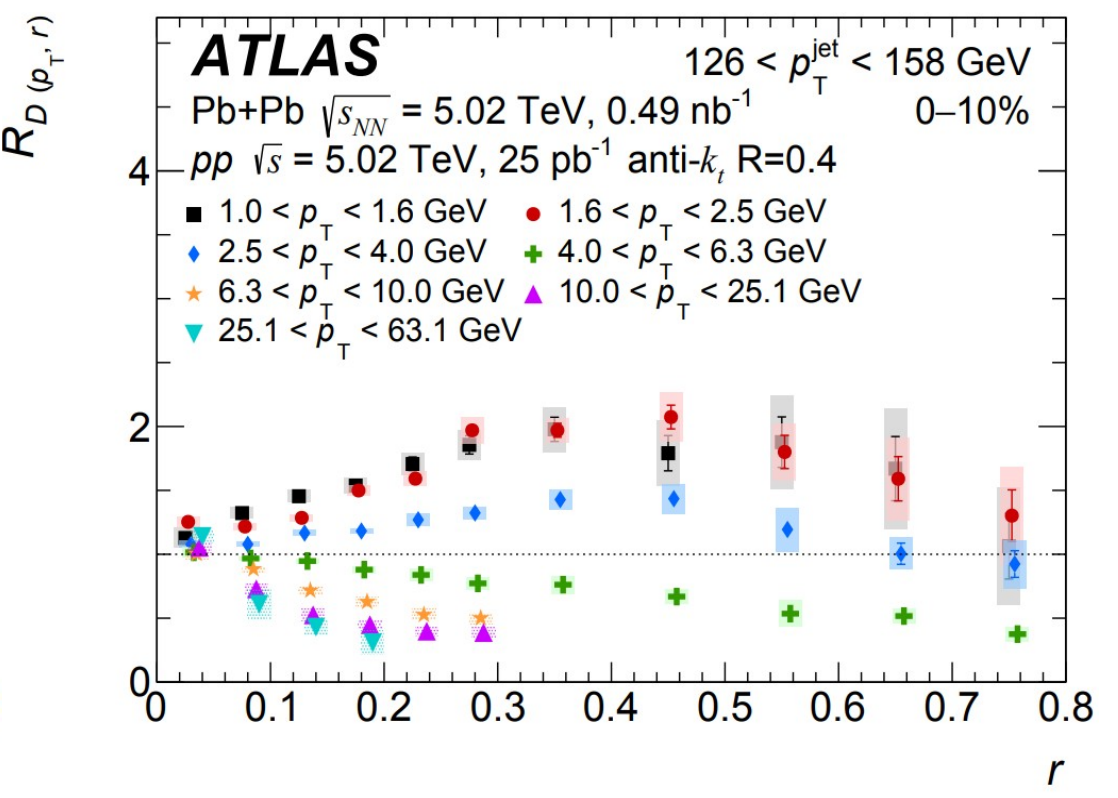}}
  \subfigure[~]{\includegraphics[width=0.54\textwidth]{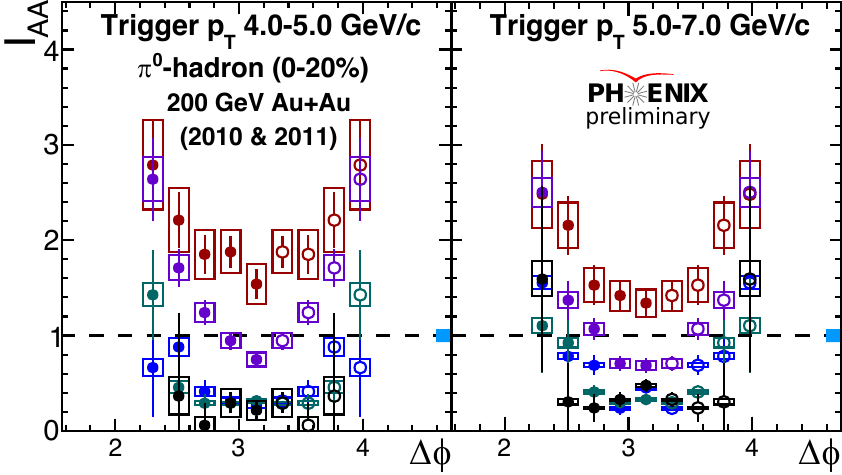}}
\caption{(a) $R_{D(p_{\rm T},r)}$ distribution as a function of angular distance $r$ for different ranges of charged-particle \pt, for 126 $<$ $p_{\rm{T, jet}}$ measured by ATLAS in central Pb--Pb collisions at \snn = 5.02~TeV. (b) $I_{AA} (\Delta \varphi) = Y^{\rm{AA}}/Y^{pp}$ in central Au--Au collisions at \snn ~=200~GeV measured by PHENIX~\cite{PHENIX_AHodges_theseProceedings}, for two trigger \pt\ ranges: 4-5 and 5-7~GeV/$c$. Red, magenta, blue, green and black colours represent associate-particle \pt\ ranges of 0.5-1~GeV/$c$, 1-2~GeV/$c$, 2-3~GeV/$c$, 3-5~GeV/$c$, 5-7~GeV/$c$, respectively.
}
\label{fig:Drad}
\end{figure}

Modification of the distribution in Pb---Pb is quantified using a ratio to $pp$, $R_{D(p_{\rm T},r)}$, as shown in Fig.~\ref{fig:Drad}(a).
An increasing enhancement of charged particles with \pt\ ~$<$~4~GeV/$c$ as a function of the angular distance $r$ from the jet axis is found, with the largest excess of particles within the jet cone. At the same time, particles with higher \pt\ are suppressed outside of the jet core. A small increase of the high-\pt\ particle yields in the core, up to $r$~=~0.05, shall be also noted.

By measuring the away-side jet correlated with a high \pt\ $\pi ^{0}$, the PHENIX collaboration also observed an increase in low momentum particle production at wide angles, consistent with a wide angle gluon radiation. 
The analysis was performed via two-particle correlation of charged hadrons with high \pt\ neutral pions, integrating the charged-particle yield in the away-side region, $Y$. Figure~\ref{fig:Drad}(b) presents the PHENIX measurement of $I_{AA} (\Delta \varphi) = Y^{\rm{AA}}/Y^{pp}$ in central Au--Au collisions at \snn ~=200~GeV for the trigger \pt\ in 4-7~GeV/$c$ range and in $|\Delta \varphi - \pi| < \pi/3$, and for different associated particle momentum range ~\cite{PHENIX_AHodges_theseProceedings}.

\section{Boson tagged jets}

Boson-tagged jets provide a controlled configuration of the initial hard scattering as photons or Z bosons do not interact strongly and are not modified by the medium. 
Moreover, at the LHC energies a high \pt\ jet sample is dominated by the quark-initiated jets. 

Recent LHC results of photon-jet correlation can be found in the following references~\cite{Sirunyan:2017qhf,Sirunyan:2018qec, Aaboud:2018anc,Aaboud:2019oac}.
Here, the latest ATLAS measurement of charged-particle per-$Z$ yields opposite in azimuth to the $Z$ boson~\cite{DPerepelitsa_theseProceedings} are presented.
Thanks to a lower level of background compared to the $\gamma$-jet case, Z bosons can be studied in a lower \pt\ range. 
Figure~\ref{fig:Flavour}(a) shows results of the charged-hadron \pt\ dependent Pb--Pb/$pp$ yield ratio, $I_{\rm{AA}}$, for two $Z$ boson \pt\ ranges of 30--60 and $>$~60~GeV/$c$.
At high hadron \pt\ ($p_{\rm{T,ch}}$) and $x_{\rm{hZ}} = p_{\rm{T,ch}}/p_{\rm{T,Z}}$, the $I_{\rm{AA}}$ is significantly suppressed and the observed trends are qualitatively similar to photon-tagged jet measurements. The Hybrid Strong/Weak Coupling model~\cite{Casalderrey-Solana:2015vaa} and $\rm {SCET_{G}}$~\cite{Chien:2015vja,Li:2019dre} (available for high $x_{\rm{hZ}}$) reproduce the experimental outcome.

\section{Flavour dependence of the energy loss}

\subsection{Gluon vs quark jets}

Due to the colour factor, gluon-seeded jets are expected to be quenched more strongly while traversing the medium, as compared to quark jets. 
The charge of a jet is sensitive to the electric charge of the initiating parton, where charge is defined as a transverse-momentum weighted sum of the electric charges of the jet constituents $q_{i}$: $Q^{\kappa}=\frac{1}{\left(p_{\mathrm{T,jet}}\right)^{\kappa}} \sum_{{i \in \rm { jet }}}{q_{i}}\left(p_{\mathrm{T}}^{i}\right)^{\kappa}$, where $\kappa$ controls sensitivity of the jet charge to low- and high-\pt\ constituents~\cite{Aad:2015cua,Sirunyan:2017tyr}. It is predicted that even in heavy-ion collisions, the charge of different flavour jets remains distinct~\cite{Li:2019dre}. And under an assumption of a larger gluon than quark energy loss, jet charge is expected to increase from $pp$ to central Pb--Pb collisions due to a larger fraction of quark jets remaining in the sample~\cite{Chen:2019gqo}. 

\begin{figure}[!htb]
\centering
  \subfigure[~]{\includegraphics[width=0.45\textwidth]{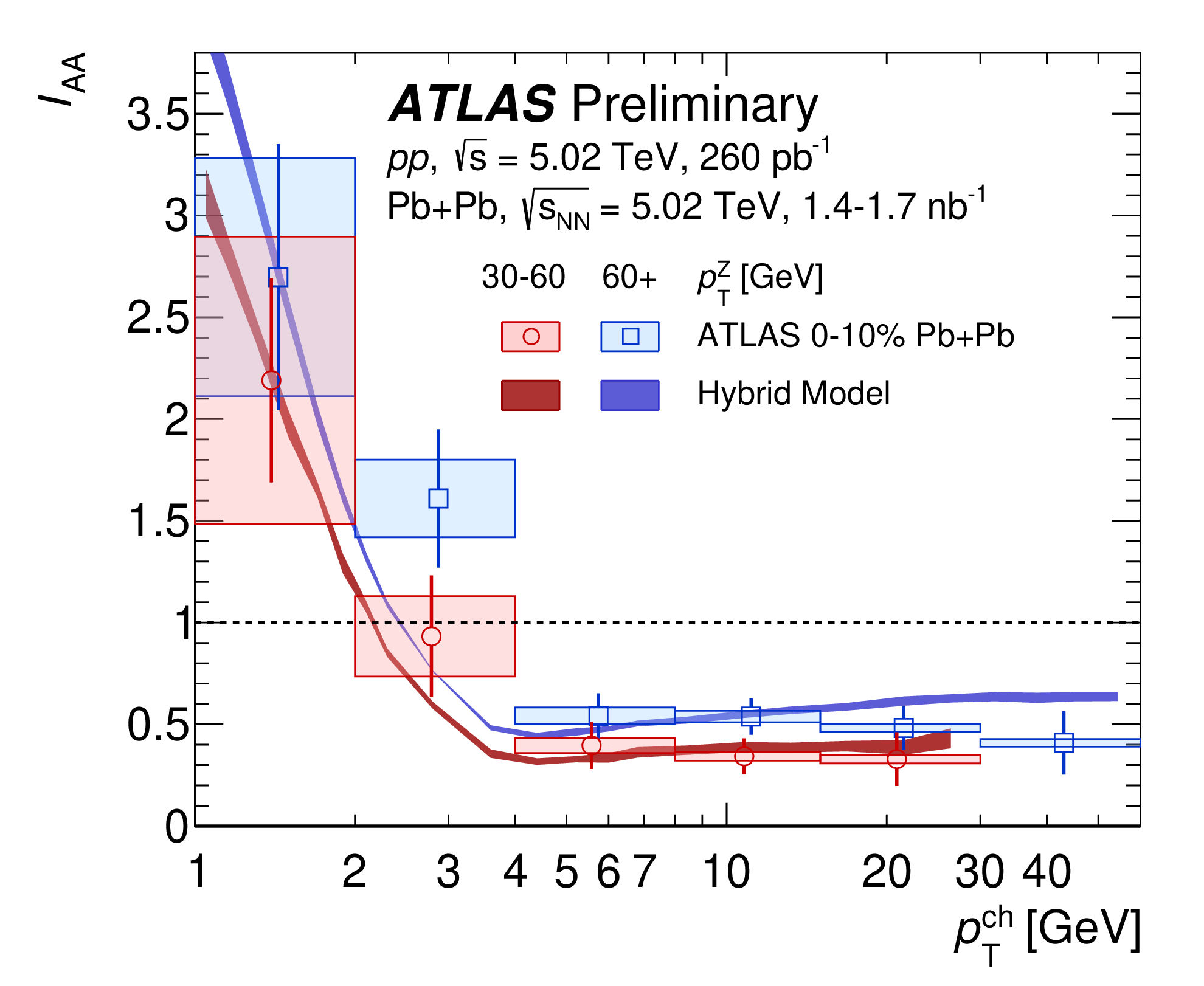}}
 \subfigure[~]{\includegraphics[width=0.54\textwidth]{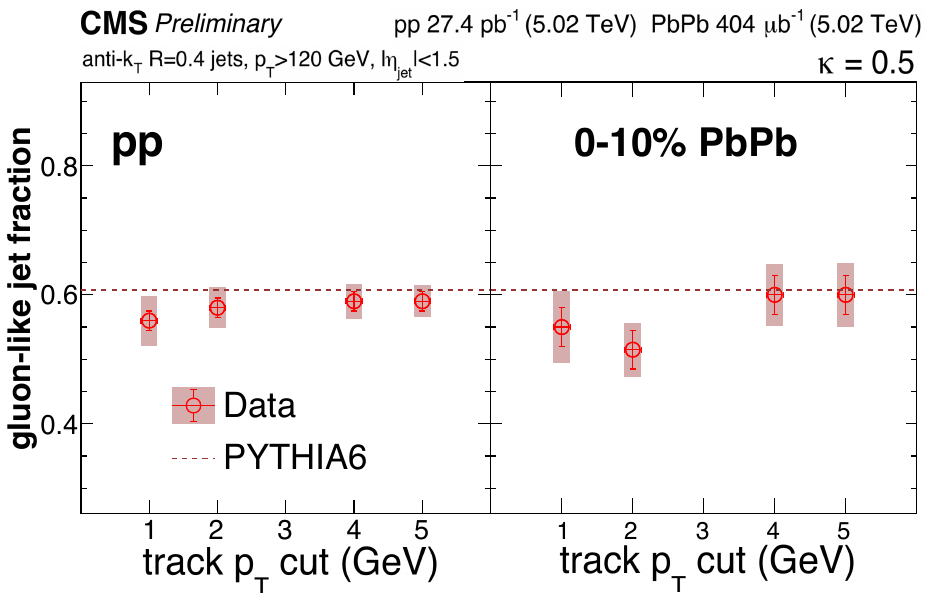}}
\caption{(a) Ratio of charged-particle yield per Z boson between central Pb--Pb and $pp$ collisions as a function of charged-hadron \pt, in two $Z$ boson \pt\ ranges, compared to the Hybrid Model predictions~\cite{Casalderrey-Solana:2015vaa}. (b) Gluon-like jet fractions in pp and 0--10\% central Pb--Pb collisions as a function of a minimum track \pt, \ptjet\ $> $ 120~GeV/$c$ and $|\eta|<$~1.5.
}
\label{fig:Flavour}
\end{figure}

Figure~\ref{fig:Flavour}(b) presents CMS results on a gluon-like jet fraction in $pp$ and central Pb--Pb collisions at \snn ~=~5.02~TeV. Jets were reconstructed using anti-$k_{T}$ algorithm with \Rpar\ ~=~0.4, \ptjet\ $> $ 120~GeV/$c$ and $|\eta|<$~1.5. The fraction of gluon jets was extracted by performing template fits to the jet charge distributions~\cite{DHangal_theseProceedings,CMS:2019exv}. 
The found gluon-like jet fraction in $pp$ collisions is about 0.6 for minimum track \pt\ $>$~1~GeV/$c$ and $\kappa$~=~0.5, consistent with PYTHIA predictions.
Contrary to the expectations, no significant modification in the gluon- and quark-like jet fractions due to jet quenching in Pb--Pb collisions w.r.t. $pp$ is observed for inclusive jets with \ptjet\ $> $ 120~GeV/$c$. In addition, an increase of the width of the jet charge, as predicted for example by PYQUEN with radiative and collisional energy loss~\cite{Lokhtin:2011qq}, is not seen in the data either.

\subsection{Heavy-flavour jets}

Studies of heavy-flavour jets provide information on the mass dependence of the parton energy loss. Due to the dead-cone effect heavier quarks are expected to lose less energy. 
But their measurements in heavy-ion collisions require high event rates and the heavy-flavour tagging is challenging due to the high background.
The available experimental results are therefore scarce. Also, not many model predictions exist so far on the market. 
Within the current experimental uncertainties, the suppression of beauty-tagged jets as measured by CMS~\cite{Chatrchyan:2013exa, Sirunyan:2018jju} is consistent with the inclusive-jet one. ALICE experiment has capability to study heavy-flavour hadrons at low \pt\ ranges. Studies of $R_{AA}$ of jets tagged with D mesons were reported in 5~$< p_{\rm{T, jet}} <$~20 GeV/$c$ in central Pb--Pb collisions~\cite{Trzeciak:2019uhd}. Strong suppression at the same level as single D meson $R_{AA}$ was observed. The suppression could not be directly compared to the inclusive jet result due to different covered \ptjet\ ranges.

Understanding the heavy-flavour jet production in $pp$ collisions is important for interpretation of heavy-ion results.
New measurements of D-meson- and b-tagged jets in $pp$ collisions at $\sqrt{s}$~=~13 and 5.02~TeV were presented~\cite{JKvapil_theseProceedings} and compared to NLO pQCD calculations of POWHEG~\cite{Frixione:2007vw} coupled to the PYTHIA generator~\cite{Sjostrand:2006za,Sjostrand:2007gs}. The measured b-jet production cross-section agrees with the central values of the predictions while the charm-jet results are consistent with the upper edge of POWHEG at low \ptjet\ as also reported before in~\cite{Acharya:2019zup}. 

\begin{figure}[!htb]
\centering
 \subfigure[~]{\includegraphics[width=0.45\textwidth]{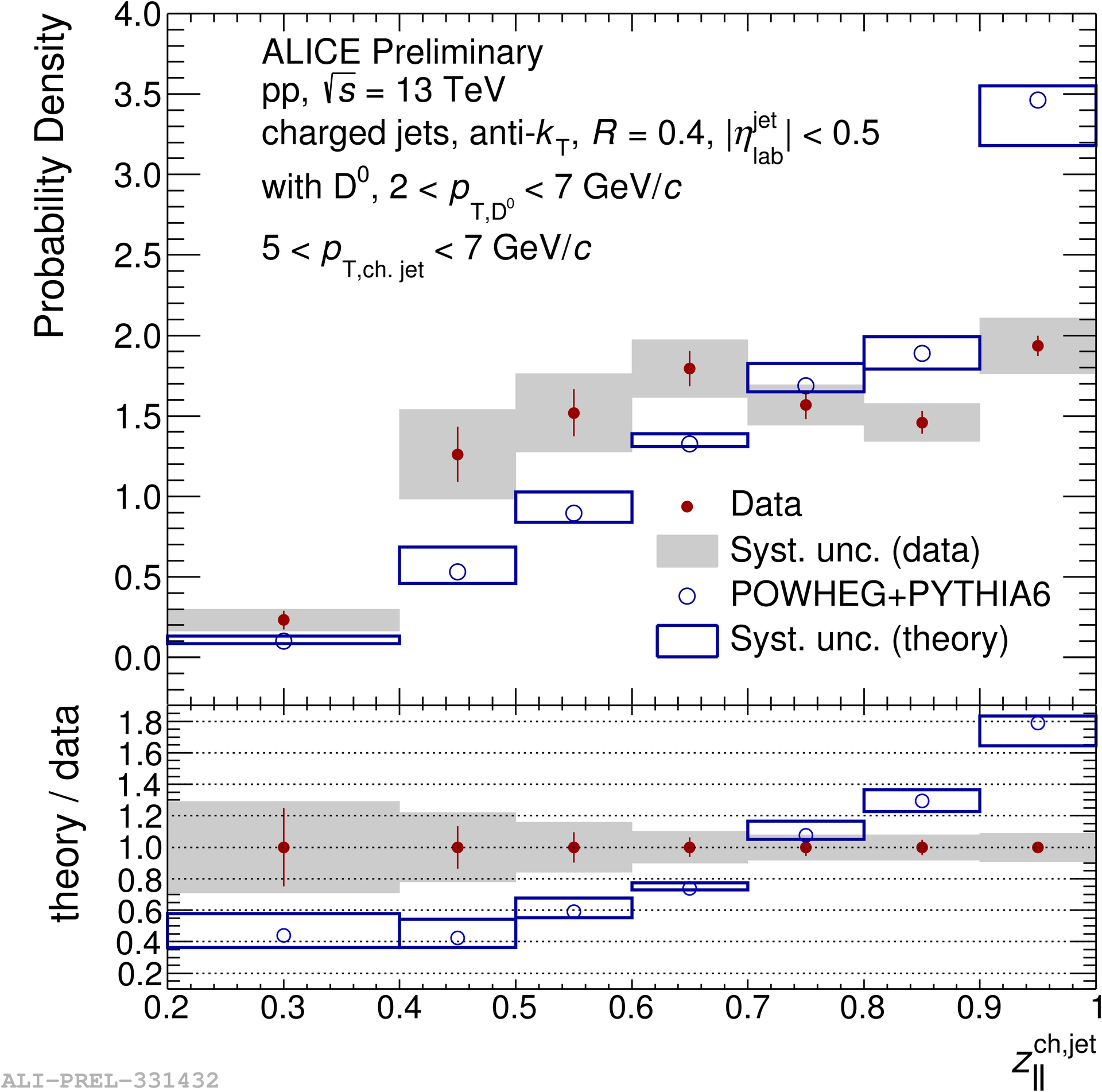}}
  \subfigure[~]{\includegraphics[width=0.54\textwidth]{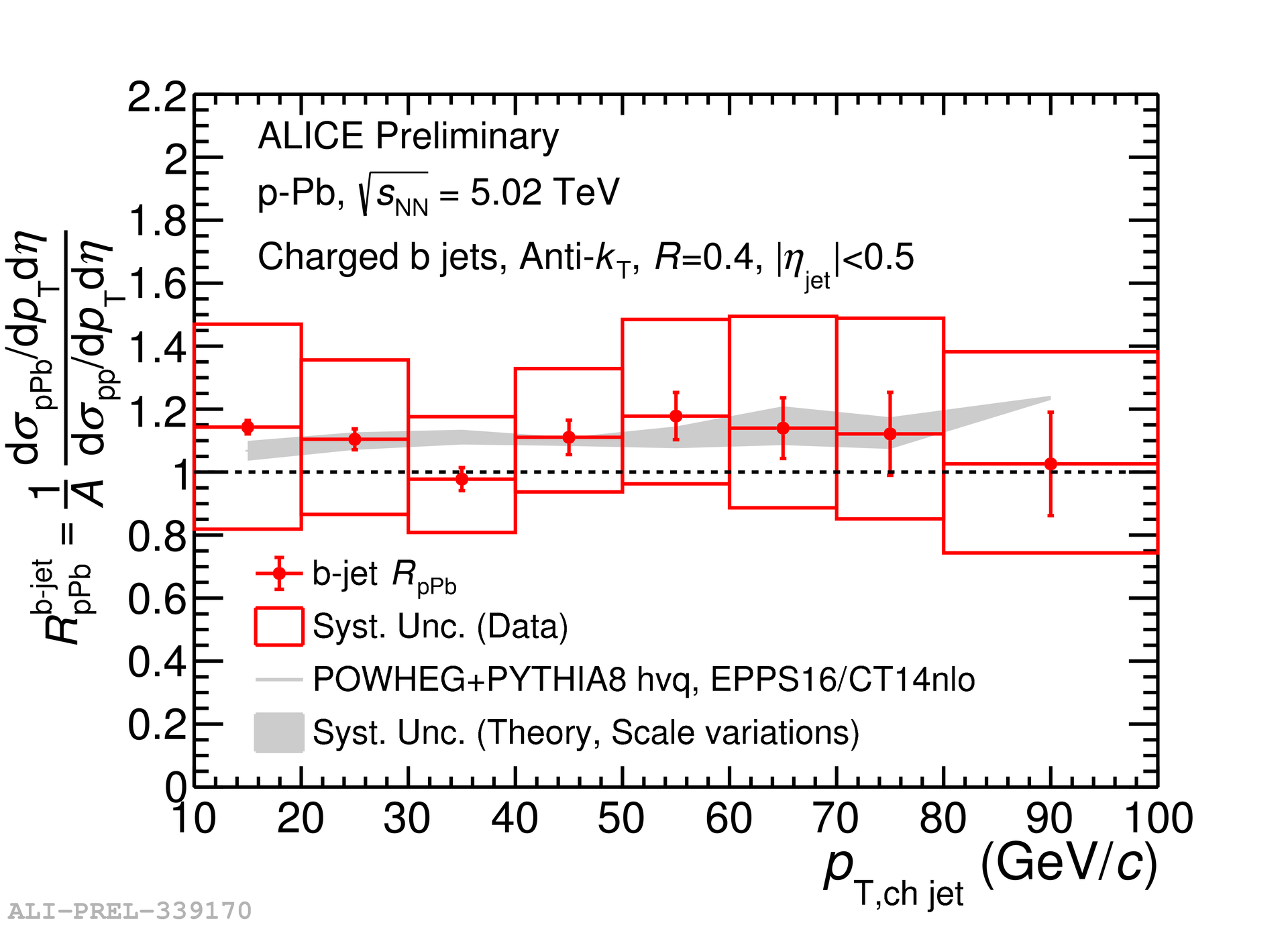}}
\caption{(a): Probability density distribution of the jet momentum fraction, $z_{||}^{\mathrm{ch}}$, carried by $\rm{D^{0}}$ meson in $pp$ collisions at $\sqrt{s}$ =~13~TeV for 5~$<$ \ptjet\ $<$~7 GeV/$c$. (b) Nuclear modification factor, $R_{pA}$, of b-tagged jets in $p$--Pb collisions at \snn\ ~=~ 5.02~TeV compared to POWHEG+PYTHIA predictions with EPPS16 nPDF.
}
\label{fig:HFjet}
\end{figure}

Furthermore, ALICE studied the fragmentation of heavy-flavour jets. The in-jet fragmentation data help to further constrain the heavy-flavour production and fragmentation in vacuum and the gluon fragmentation functions~\cite{Anderle:2017cgl}.
New measurements of the jet momentum fraction carried by the D meson or $\Lambda_{C}$ baryon, $z_{||}^{\mathrm{ch}}$:
\begin{equation}
z_{||}^{\mathrm{ch}}=\frac{\vec{p}_{\mathrm{D(\Lambda_{C})}} \cdot \vec{p}_{\mathrm{ch, jet }}}{\vec{p}_{\mathrm{ch, jet}} \cdot \vec{{p}}_{\mathrm{ch,jet}}} ,
\label{eg:z}
\end{equation}
in $pp$ collisions at $\sqrt{s}$~=~5 and 13~TeV were presented~\cite{JKvapil_theseProceedings}.
In case of the D-tagged jets, a hint of softer fragmentation than the POWHEG+PYTHIA predicts was observed at low \ptjet\ (5-7~GeV/$c$), as can be seen in Fig.~\ref{fig:HFjet}(a).

Similarly as in the case of the inclusive jet production in $p$A collisions~\cite{Adare:2015gla,Khachatryan:2016xdg,Chatrchyan:2014hqa,ATLAS:2014cpa,Adam:2015hoa}, the heavy-flavour jet production is not modified w.r.t. $pp$ collisions~\cite{JKvapil_theseProceedings}, which is consistent with an absence of a strong final state suppression. $R_{pA}$ of heavy-flavour-electron-(HFE) and b-tagged jets is consistent with unity and with expectation from the POWHEG+PYTHIA with EPPS16 nPDFs~\cite{Eskola:2016oht}, as presented in Fig.~\ref{fig:HFjet}(b). The HFE-tagged jet production was analysed for different \Rpar\ values, \Rpar\ ~=~0.3, 0.4 and 0.6, in both $pp$ and $p$--Pb collisions. No dependence of the $R_{pA}$ on $R$ and no modification of the jet shape in $p$--Pb collisions, studied via $\sigma(R=0.3)/\sigma(R=0.6)$ ratio, was found~\cite{JKvapil_theseProceedings, SSakai_poster}.

\section{Summary}

Thanks to the recent LHC and RHIC studies of the jet and high \pt\ hadron production in heavy-ion collisions, a  more coherent picture of the parton in-medium energy loss starts to emerge. 
It has been observed that the lost energy is redistributed to large angles and soft particle multiplicities. Extensive R-dependent jet measurements were also performed.
These new experimental results give better constrains on the jet quenching models, that should improve our understanding of the underlying parton energy loss mechanism and the medium response. 

From the experimental side, it would be interesting to extend the current jet studies to large \Rpar\ in a wider \ptjet\ range, particularly in central collisions. 
Concerning the heavy-flavour tagged jets, the LHC Pb--Pb data from year 2018 have a potential to provide more precise results on charm and beauty jets, also in \pt\ intervals overlapping with the inclusive jet measurements. Moreover, it is important to perform complementary jet studies in heavy-ion collisions both at RHIC and LHC, varying in this way the medium temperature and probing different quark/gluon composition.

\section*{Acknowlegment}
This work was supported by the grant CZ$.02.1.01/0.0/0.0/16\_013/001569$ (Brookhaven National Laboratory - participation of the Czech Republic) of Ministry of Education, Youth and Sports of the Czech Republic.

\bibliographystyle{h-elsevier}
\bibliography{bibQM19}

\end{document}